# Static Rashba Effect by Surface Reconstruction and Photon Recycling in the Dynamic Indirect Gap of APbBr$_3$ (A = Cs, CH$_3$NH$_3$) Single Crystals

Hongsun Ryu,[†] Dae Young Park,[§] Kyle M. McCall, [∥,‡,¶] Hye Ryung Byun,[†] Yongjun Lee,[§] Tae Jung Kim,[⊥] Mun Seok Jeong,[§] Jeongyong Kim, [§] Mercouri G. Kanatzidis, [∥] and Joon I. Jang*,[†]

[†] Department of Physics, Sogang University, Seoul, 04107, South Korea
[§] Department of Energy Science, Sungkyunkwan University, Suwon, 16419, South Korea
[∥] Department of Chemistry, Northwestern University, Evanston, Illinois 60208, USA
[‡] Laboratory of Inorganic Chemistry, Department of Chemistry and Applied Biosciences, ETH Zurich, 8093 Zurich, Switzerland
[¶] Laboratory for Thin Films and Photovoltaics, Empa - Swiss Federal Laboratories for Materials Science and Technology, CH-8600 Dübendorf, Switzerland
[⊥] Department of Physics, Kyung Hee University, Seoul, 02447, South Korea

**ABSTRACT:** Recently, halide perovskites have gained significant attention from the perspective of efficient spintronics owing to Rashba effect. This effect occurs as a consequence of strong spin-orbit coupling under noncentrosymmetric environment, which can be dynamic and/or static. However, there exist intense debates on the origin of broken inversion symmetry since the halide perovskites typically crystallize into a centrosymmetric structure. In order to clarify the issue, we examine both dynamic and static effects in the all-inorganic CsPbBr$_3$ and organic-inorganic CH$_3$NH$_3$PbBr$_3$ (MAPbBr$_3$) perovskite single crystals by employing temperature- and polarization-dependent photoluminescence excitation spectroscopy. The perovskite single crystals manifest the dynamic effect by photon recycling in the indirect Rashba gap, causing dual peaks in the photoluminescence. But the effect vanishes in CsPbBr$_3$ at low temperatures (< 50 K), accompanied by a striking color change of the crystal, arising presumably from lower degrees of freedom for inversion symmetry breaking associated with the thermal motion of the spherical Cs cation, compared with the polar MA cation in MAPbBr$_3$. We also show that static Rashba effect occurs only in MAPbBr$_3$ below 90 K due to surface reconstruction via MA-cation ordering, which likely extends across a few layers from the crystal surface to the interior. We further demonstrate that this static Rashba effect can be completely suppressed upon surface treatment with poly methyl methacrylate (PMMA) coating. We believe that our results provide a rationale for the Rashba effects in halide perovskites.

## INTRODUCTION

Halide perovskites have emerged as "miracle" materials for optoelectronic applications because of their superior optical, electronic, and structural properties such as high absorption coefficient,[1] defect tolerance,[2] bandgap tunability,[3–5] and easy preparation methods.[6–8] For instance, organic-inorganic halide perovskites (OIHPs) have been investigated in the form of thin films for highly efficient light harvesters, as they have a long carrier lifetime and a long diffusion length.[6,9,10] On the other hand, all-inorganic halide perovskites (AIHPs) have been prepared into nanocrystals for highly efficient light emitters, as they possess high photoluminescence (PL) quantum yields,[11,12] relatively short PL lifetimes,[13] and better stability.[14] But the nanocrystals are not quite ideal for studying intrinsic properties of halide perovskites because they are sensitive to surface defects with a high surface to volume ratio[15] and thin films are influenced by substrates[16] and grain boundaries.[17] Obviously, a single crystal seems more suitable for analyzing pristine optical properties of these materials with minimal extrinsic effects. Intriguingly, however, the optical behavior of a perovskite single crystal is complicated by an additional feature in the low-energy tail of PL emission[5,18–20] arising from cascade of emission and reabsorption[21] in the bulk, namely photon recycling.[22] The resulting dual-peak feature often led to the incorrect concept of above-bandgap-emission,[23–26] whereas strong reabsorption renders a typical carrier dynamics unreliable. In fact, there were several attempts to explain nontrivial carrier dynamics in perovskite single crystals by including fast diffusion of carriers into the interior,[27–29] but there exists a wide variance in dynamic parameters that are strongly model dependent.

Photon recycling requires a subgap state to cause dual-gap emission of the PL in halide perovskites. Although typical Urbach tailing can serve as a subgap state, an alternative mechanism was recently proposed to account for the dual nature in terms of additional indirect transition allowed by dynamic Rashba effect.[18] When a nonzero electric field is generated by inversion symmetry breaking, the spin of a free charge couples to an effective magnetic field by strong spin-orbit coupling (SOC), causing spin-dependent energy band splitting to form an indirect gap slightly below the direct band edge. Surprisingly, despite the presence of an inversion center, evidence for Rashba splitting has been reported in halide perovskites by computational simulations,[30–32] circular photo-galvanic effect,[33] spin-polarization dependence,[18] optical Hanle effect,[34] and so on. Currently, there are various views on the origin for broken inversion symmetry, which are categorized into dynamic effects[18,33,35–37] (arising from dynamic lattice distortion on a 100 fs time scale[38,39]) and static effects (arising from static inversion symmetry breaking),[34,39–41] but the issue remains controversial.

In this paper, we aim to identify and even control the underlying mechanisms for inversion symmetry breaking that cause both dynamic and static Rashba effects in high-quality single crystals based on a series of precision spectroscopic measurements, which include nonlinear optical multiphoton absorption and harmonic generation. First, we precisely pinpointed the location of the indirect gap by fine-scale photoluminescence excitation (PLE) spectroscopy over a wide temperature range from 10 K to 300 K. The degree of Rashba splitting was assessed by the energy difference between the direct gap and the indirect gap,[18] which shrinks with decreasing temperature. While dynamic Rashba effect persists in MAPbBr$_3$, we confirmed that it essentially vanishes in CsPbBr$_3$ below 50 K, which is also evidenced by the recovery of the color purity of the crystal. This in turn causes significant increase in the PL brightness via suppression of photon recycling in the dynamic Rashba gap. We found that dynamic Rashba effect is critically dependent on the type of the A-site cation (Cs or MA) and the origin for dynamic inversion symmetry breaking arises most likely from thermal motion of the cation in the perovskite cage. Then, we showed that static Rashba effect develops only in MAPbBr$_3$ at low temperatures (< 90 K) based on temperature- and polarization-dependent PL spectroscopy. The degree of circular polarization in MAPbBr$_3$ increases at lower temperature and can be as high as ~3% at 10 K. The origin for static inversion symmetry breaking is local reconstruction of surface via MA-cation ordering, which is most significant across a few layers from the crystal surface to the interior as corroborated by nonlinear optical measurements. We further demonstrate that this static effect in the OIHP can be completely quenched by a simple surface treatment using poly methyl methacrylate (PMMA) coating. We emphasize that our results do not show any evidence of Rashba splitting by extrinsic effects due to defects or imperfections, and therefore, provide a rationale for both dynamic and static Rashba effects in this important class of materials in pristine quality. We believe that our fundamental results have implications in the role of halide perovskites in the emerging field of spin-optoelectronics, which is combination of spintronics and optoelectronics.[34,42]

## RESULTS AND DISCUSSION

Our CsPbBr$_3$ and MAPbBr$_3$ single crystals were synthesized by the Bridgman[14] and inverse temperature crystallization (ITC) methods,[43] respectively. The sample preparation procedure and X-ray diffraction results (Figure S1) are detailed in the Supporting Information. Figures 1a and 1d show the absorption (black) and the PL spectra from CsPbBr$_3$ and MAPbBr$_3$, respectively, obtained at 300 K under reflection geometry (R mode, blue) and transmission geometry (T mode, red). The corresponding excitation wavelength ($\lambda_{ex}$) for one-photon absorption (1PA) was 500 nm. The intensity dependence of the PL clearly indicates that it arises from radiative recombination of free carriers at room temperature (Figure S2). The peak position ($\lambda_d$) of the blue trace coincides with the absorption edge, which corresponds to the wavelength for the fundamental direct gap of each single crystal; at low temperatures where excitons are dominant, $\lambda_d$ corresponds to the wavelength for the exciton PL, i.e., the optical gap that is slightly below the direct gap by exciton binding energy.

Under the T mode, however, a completely different PL (red trace) was obtained from each sample in terms of both peak position and brightness. This drastic change in the PL arises from effective reabsorption of the main PL while traveling a macroscopic distance before exiting the sample (thickness ~2 mm). We attribute the origin for this subgap absorption to the indirect Rashba gap arising from dynamic inversion symmetry breaking in accordance with Ref. [[33]] and as further explained below through its temperature dependence (Figure 2). Hence, the high-energy onset of the red trace was assigned to $\lambda_i$ that corresponds to the wavelength for the indirect gap formed slightly below the direct gap. This implies that the single crystal is truly transparent above $\lambda_i$ (below the Rashba gap). It is interesting that the dynamic Rashba gap can be measured by time-integrated absorption or PL experiments, because dynamic inversion symmetry breaking is always present throughout random thermal distortion of the perovskite cage. However, we found that this dynamic effect can be discerned by neither circular polarization spectroscopy nor second harmonic generation (SHG) as they have directional dependence, which would be averaged out for the random thermal motion. It is noteworthy that the main PL (blue trace) is spectrally asymmetric because of low-energy tailing, which indeed coincides with the red trace. This dual-peak nature obtained under the R-mode therefore consists of a symmetric band-to-band transition and the additional PL resulting from photon-recycling via successive emission and reabsorption of the main PL directed into the interior of the sample. As described later, the low-energy-peak intensity strongly depends on the sample thickness (Figure 5). This indicates that the secondary peak is not simple indirect emission but ambient PL after photon recycling through the dynamic Rashba gap.

In order to determine the precise locations for $\lambda_d$ and $\lambda_i$, we performed fine-scale PLE spectroscopy on our single crystals. Figures 1b and 1e show the series of the PL spectra from CsPbBr$_3$ and MAPbBr$_3$, respectively, plotted on a semi-logarithmic (semi-log) scale when $\lambda_{ex}$ was varied from $\lambda_i$ to $\lambda_d$. The sharp peak on top of each PL spectrum corresponds to the scattered laser light registered as a peak by our detection system, where its intensity was cut down by more than a factor of 100 with orthogonal polarization control. Interestingly, both single crystals yield a weak indirect PL when $\lambda_{ex}$ is resonant with the Rashba gap, i.e., $\lambda_{ex} = \lambda_i$. When $\lambda_{ex}$ was varied from the Rashba onset to the direct gap, the PL intensity steeply increases and its peak position blueshifts towards the direct gap, until the peak position saturates into a constant wavelength, which is $\lambda_d$. In Figures 1c and 1f, we plot the corresponding PL peak position (black spheres) and the PL intensity (blue spheres) as a function of $\lambda_{ex}$ as determined from Figures 1b and 1e. The values for $\lambda_d$ are 527 nm (~2.35 eV) and 539 nm (~2.30 eV) for CsPbBr$_3$ and MAPbBr$_3$, respectively, at 300 K. These bandgap wavelengths are consistent with values for $\lambda_d$ shown in Figures 1a and 1d. The values for $\lambda_i$ were determined from the onset $\lambda_{ex}$, causing the weak indirect PL, which are 551 nm (2.25 eV) and 567 nm (~2.19 eV) for CsPbBr$_3$ and MAPbBr$_3$, respectively, at 300 K. These indirect-gap wavelengths are consistent with the values for $\lambda_i$ shown in Figures 1a and 1d. This implies that the PL spectra using the R and T modes can be utilized for estimating $\lambda_d$ and $\lambda_i$. Based on the difference between $\lambda_i$ and $\lambda_d$, we conclude that Rashba splitting is about 10% larger for the OIHP single crystal.

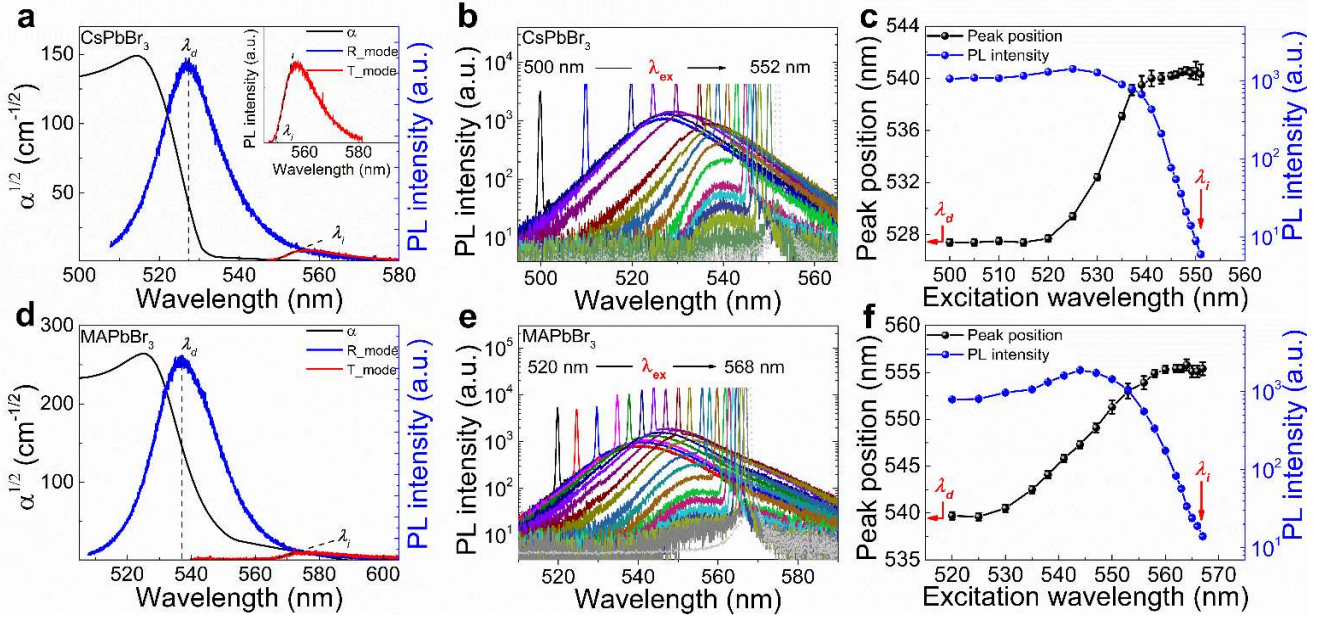

**Figure 1.** Absorption (black) and PL spectra from (a) CsPbBr$_3$ and (d) MAPbBr$_3$ obtained under R mode (blue) and T mode (red) at 300 K when excited at $\lambda_{ex}$ = 500 nm. The wavelengths for the indirect and direct gaps are denoted by $\lambda_i$ and $\lambda_d$, respectively. Semi-log plot of the series of room-temperature PL spectra from (b) CsPbBr$_3$ and (e) MAPbBr$_3$ obtained using PLE spectroscopy. The PL peak position (black spheres) and the PL intensity (blue spheres) as a function of $\lambda_{ex}$ at 300 K determined by the PLE data.

In order to study the temperature dependence of dynamic Rashba effect, we measured the PL as a function of temperature from 300 K to 10 K (Figures S3–S5). For instance, in Figures 2a and 2d, we plot the PL obtained from CsPbBr$_3$ and MAPbBr$_3$, respectively, at 10 K under the R mode (blue) and the T mode (red). At this low temperature, the nature of the PL is essentially excitonic (X) since exciton binding energy is much larger than thermal energy (~0.9 meV).[44,45] However, we found that MAPbBr$_3$ exhibits a small but measurable free-carrier peak (FC) even at 10 K, which requires further investigation in the future. In both single crystals, the so-called P-band emission (P) is dominant, which occurs near the polariton bottleneck being slightly below the X line as a consequence of inelastic exciton-exciton scattering.[46,47] Note that the effect of reabsorption is significantly suppressed in CsPbBr$_3$, as indicated by the coincidence between the blue and the red traces across the Rashba gap in Figure 2a. Interestingly, this is accompanied by the striking color change of the crystal from pale orange (300 K) to transparent yellow (10 K) as shown in the inset. The color change arises from the recovery of the true color of the sample without additional absorption of ambient light by the indirect Rashba gap. In contrast, MAPbBr$_3$ does not exhibit a significant color change at 10 K, implying that dynamic Rashba effect persists at this low temperature. Therefore, there exists reabsorption of the PL in MAPbBr$_3$, causing a clear difference in the blue and red traces in Figure 2d. The estimated values for $\lambda_d$ and $\lambda_i$ are 536.9 nm and 537.6 nm in CsPbBr$_3$ and 551.8 nm and 562.1 nm in MAPbBr$_3$, respectively.

In Figures 2b and 2e, we plot $\lambda_d$ (blue squares) and $\lambda_i$ (red squares) of CsPbBr$_3$ and MAPbBr$_3$, respectively, as a function of temperature based on the PL spectra shown in Figures 2a, 2d, S3, and S4. As mentioned, the Rashba gap essentially vanishes in CsPbBr$_3$ below 50 K ($\lambda_i - \lambda_d < 1$ nm), whereas in MAPbBr$_3$ it remains almost a constant ($\lambda_i - \lambda_d \sim 10$ nm) below the dashed line (~140 K), where a structural phase transition[26] occurs from the tetragonal phase to the orthorhombic phase. A sudden change in $\lambda_i - \lambda_d$ across the dashed line seems to indicate that dynamic inversion symmetry breaking is less pronounced in the orthorhombic structure having a lower symmetry. Except for this complication arising from the phase transition in MAPbBr$_3$, both single crystals undergo a monotonic bandgap redshift as temperature decreases, which is typical for halide perovskites.[18,48] But it should be noted that the temperature dependence of the indirect gap ($\lambda_i$) is not related to any bandgap-shift model, such as Varshni equation,[49] as it solely reflects the degree of dynamic Rashba effect. A drastically different behavior observed at low temperatures shows that dynamic inversion symmetry breaking is contingent upon the type of the A-site cation (Cs or MA) and arises most likely from thermal motion of the cation in the perovskite cage. It is also evident that dynamic Rashba effect is more prominent in MAPbBr$_3$ due to the polar character of the MA cation, compared with the much more symmetric Cs cation in CsPbBr$_3$.

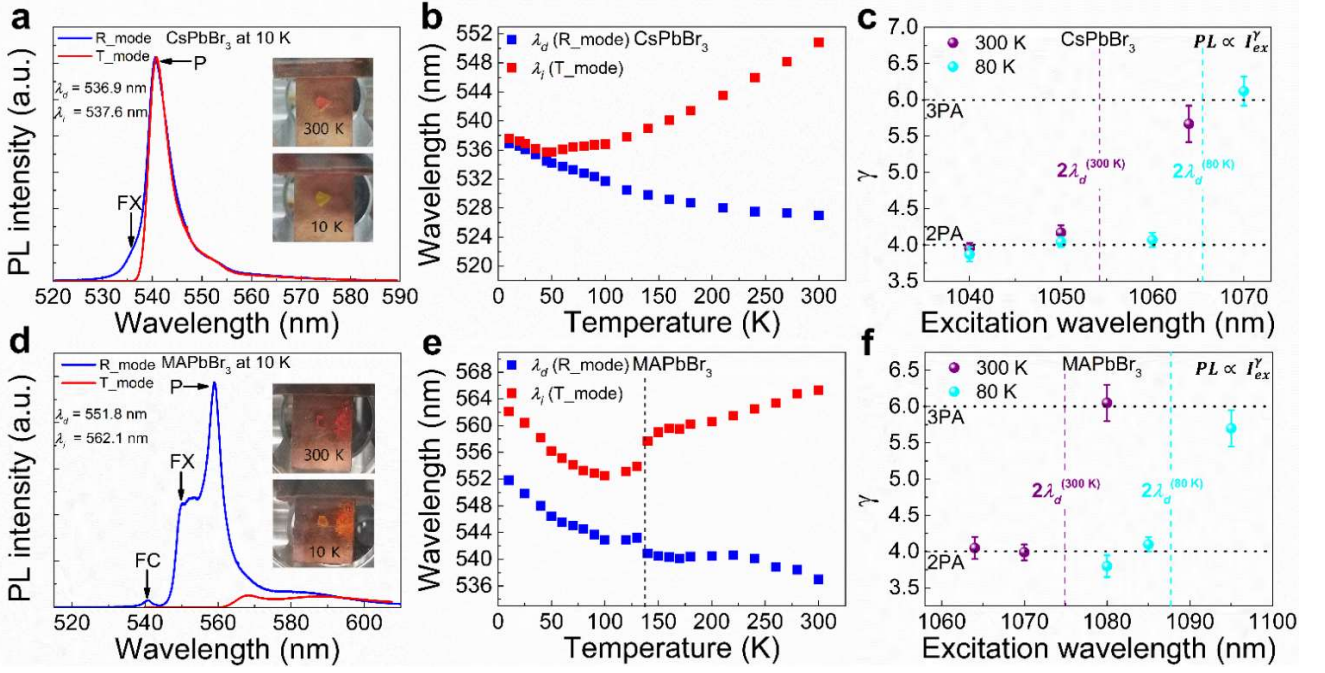

**Figure 2.** PL spectra from (a) CsPbBr$_3$ and (d) MAPbBr$_3$ at 10 K obtained under the R mode (blue) and the T mode (red). The inset shows the representative photographs of the single crystals at 300 K and 10 K. The apparent color change in CsPbBr$_3$ is associated with quenching of dynamic Rashba effect. Temperature-dependent $\lambda_d$ (blue squares) and $\lambda_i$ (red squares) of (b) CsPbBr$_3$ and (e) MAPbBr$_3$ at $\lambda_{ex}$ = 500 nm. The dashed line indicates the phase-transition temperature for MAPbBr$_3$ around 140 K. Power exponent ($\gamma$) of the intensity-dependent PL vs. $\lambda_{ex}$ for (c) CsPbBr$_3$ and (f) MAPbBr$_3$ at 80 K and 300 K, where the boundaries for 2PA and 3PA are indicated by the cyan dashed line (80 K) and the purple dashed line (300 K), respectively. The $\gamma$ values of ~4 for 2PA and ~6 for 3PA indicate that the PL arises from carrier recombination at high temperatures (> 80 K).

We performed nonlinear optical multiphoton absorption spectroscopy across the boundary between two-photon absorption (2PA) and three-photon absorption (3PA) for the perovskite crystals and examined the excitation power dependences at two representative temperatures of 300 K and 80 K (see for example Figures S6 and S7 for CsPbBr$_3$). Here we emphasize that the multiphoton absorption order must be defined in terms of $\lambda_d$, not $\lambda_i$.[23] The purple and cyan dashed lines in Figures 2c and 2f correspond to the 2PA-3PA boundaries for CsPbBr$_3$ and MAPbBr$_3$ at 300 K and 80 K, respectively. The purple spheres (300 K) and cyan spheres (80 K) correspond to the power exponents ($\gamma$'s) for the selected excitation wavelengths, indicating that the PL arises from radiative recombination of free carriers as $\gamma$ values are about 4 for 2PA (left side of the dashed lines) and 6 for 3PA (right side of the dashed lines), respectively. We found that a minor deviation from the precise power law is due to the saturation effect of the carriers during the pulse time (30 ps).[47]

Under strong SOC and inversion symmetry breaking, the spin degeneracy of the valence band (VB) and the conduction band (CB) can be removed. In this situation, selective optical excitation of a specific spin band can be realized by using right/left circularly polarized light (RCP/LCP, $\sigma^-/\sigma^+$). Splitting of the VB (CB) is mainly contributed by hybridization of Br 4$p$ and Pb 6$s$ orbitals (Pb 6$p$ orbitals), causing a larger splitting in the CB than the VB as schematically depicted in the inset of Figure 3a.[32,50] Therefore, the presence of Rashba splitting can be confirmed by measuring the degree of circular polarization (DOCP) given by

$$\text{DOCP (\%)} = \left|\frac{\text{PL}(\sigma^+) - \text{PL}(\sigma^-)}{\text{PL}(\sigma^+) + \text{PL}(\sigma^-)}\right| \times 100 \quad (1)$$

where PL($\sigma^-$) and PL($\sigma^+$) are the measured PL intensities under RCP and LCP excitation, respectively. Considering a fast spin relaxation time due to strong SOC in halide perovskites, the DOCP would be most visible when employed to the case for time-resolved excitation,[51,52] but can be certainly used for our time-integrated PL spectroscopy. However, it should be noted that dynamic Rashba effect would yield DOCP = 0% in the time-integrated regime, since the orientation of spins is random under thermal distortion of the lattice. Therefore, in our experimental configuration, a nonzero DOCP would be a smoking gun for static Rashba effect, where the origin for inversion symmetry breaking is systematic, not thermal.

Figure 3a illustrates our experimental setup for polarization-dependent PL spectroscopy to assess the DOCP of the perovskites using a pair of quarter waveplates (QWP's) (see the Supporting Information for details). Compared with a typical setup using a beam splitter, we collected the PL signal under scattering geometry ($\theta \sim 20°$) in order to entirely eliminate the beam-splitting efficiency that depends on polarization as well as wavelength. In Figure 3b we plot the PL spectra from CsPbBr$_3$ using $\lambda_{ex}$ = 532 nm being ~10 nm above the exciton PL line at 10 K, where the trace in the upper (lower) panel corresponds to the PL excited by $\sigma^+(\sigma^-)$ polarization and the red (black) trace being the PL detected using $\sigma^+(\sigma^-)$ polarization. In other words, the red (black) PL in the upper panel corresponds to the parallel (orthogonal) polarization scheme and the color code reverses in the lower panel. Ideally, the PL pair with different colors in the upper and lower panels should be a mirror image each other. Throughout careful analyses on the series of the PL data averaged over laser fluctuation we confirmed that the DOCP is zero in our AIHP single crystal over the entire temperature range (Figure S8). Hence, there is no static inversion symmetry breaking in CsPbBr$_3$. In fact, we did not find any evidence for extrinsic symmetry-breaking sources such as defects or impurities to cause measurable static Rashba effect, thereby signifying the quality of our specimen.

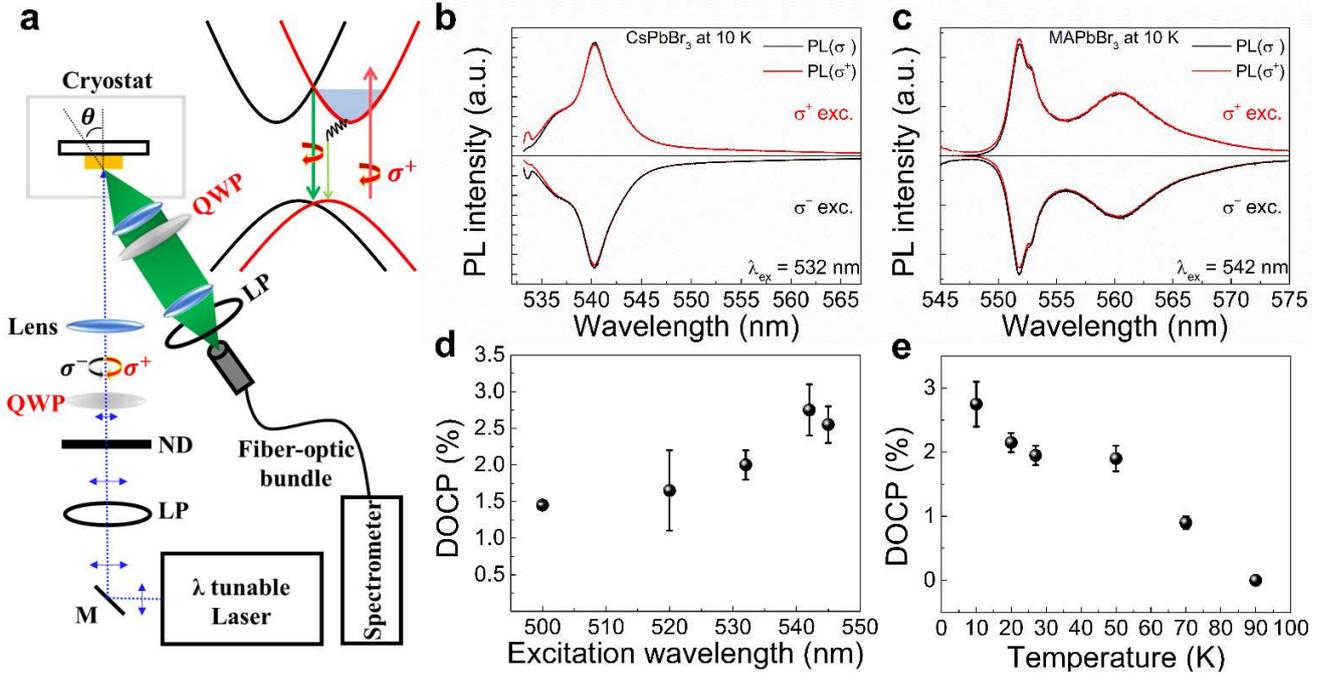

**Figure 3.** (a) Experimental setup for the DOCP measurement using scattering geometry ($\theta \sim 20°$) to remove the polarization dependence of a beam splitter, where M is a mirror, ND is a neutral-density filter, QWP's are quarter waveplates, and LP's are linear polarizers. The inset illustrates the schematic of the Rashba-split conduction and valence bands with spin-dependent selective optical transition for circularly polarized light ($\sigma^+$) near the bandgap ($E_g$). Right/left circularly polarized PL($\sigma^-$)/PL($\sigma^+$) under $\sigma^-/\sigma^+$ excitation from (b) CsPbBr$_3$ and (c) MAPbBr$_3$ at 10 K. Only MAPbBr$_3$ exhibits a finite DOCP near the direct-gap peak (552 nm). (d) $\lambda_{ex}$-dependent DOCP of MAPbBr$_3$ at 10 K. (e) Temperature-dependent DOCP of MAPbBr$_3$ by tuning $\lambda_{ex}$ being ~10 nm above the exciton PL line.

However, our OIHP single crystal exhibits a small but finite DOCP (~3%) near the exciton PL at 10 K as shown in Figure 3c, in which $\lambda_{ex}$ was tuned to 542 nm in order to ensure the same excitation condition, i.e., being ~10 nm above the main exciton PL line. We emphasize that DOCP should be assessed at the direct transition, as it is zero for indirect transition involving incoherent phonon scattering. Most importantly, we confirmed that this finite DOCP is always reproducible over any excitation spot on the sample, clearly demonstrating the presence of static Rashba effect. This static effect can be more clearly seen via the PL contrast between the parallel and orthogonal polarization as shown in Figure S9, which plots |PL($\sigma^+$) − PL($\sigma^-$)|. We also measured DOCP of MAPbBr$_3$ as a function of $\lambda_{ex}$ and temperature as well (Figures S10 and S11). The dots in Figure 3d correspond to the DOCP values at 10 K for several excitation wavelengths, showing that DOCP increases up to nearly 3% when $\lambda_{ex}$ was tuned towards the main PL. This effect can be understood in terms of an available phase space for spin flip and indeed observed in the recent work.[18] Figure 3e depicts the DOCP values (dots) of MAPbBr$_3$ as a function of temperature. DOCP assumes the maximum value of ~3% at the lowest temperature but monotonically decreases with increasing temperature and vanishes at 90 K. Therefore, we conclude that the OIHP exhibits both dynamic and static Rashba effects, where the latter is present at low temperatures < 90 K.

The origin for the static inversion symmetry breaking is currently being debated by several candidates: (i) ferroelectric structure,[40] (ii) anharmonic thermal fluctuation between organic cation and PbBr$_6$,[18,33,35–37] (iii) MA dipole alignment at low temperature,[34] (iv) structural surface relaxation,[53] and so on. We measured the SHG response from the single crystals at 10 K at $\lambda_{ex}$ = 1500 nm, as SHG is sensitive to static inversion symmetry breaking in the crystalline phase. More specifically, our picosecond excitation setup is ideal for sensing bulk SHG only,[54] in which a 30-ps pulse is too weak in input intensity to cause surface SHG arising from the single boundary layer, compared with a typical femtosecond setup. As predicted, CsPbBr$_3$ does not exhibit any SHG (Figure 4a), consistent with zero DOCP. Intriguingly, however, it turns out that MAPbBr$_3$ does not yield any measurable SHG response (black) either as shown in Figure 4b. The absence of bulk SHG was further confirmed by the strong third harmonic generation (THG) response (red), which is a higher-order nonlinear optical process and generally weaker than SHG by orders of magnitude in typical noncentrosymmetric crystals. Therefore, the origin for static inversion symmetry breaking in the OIHP single crystal should be essentially surface induced.

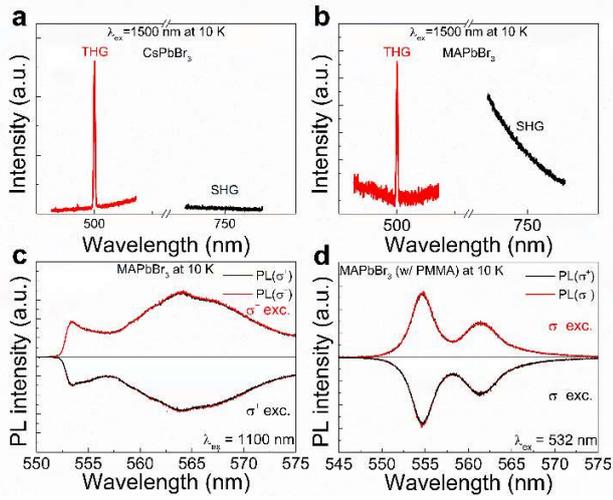

**Figure 4.** SHG (black) and THG (red) responses from (a) CsPbBr$_3$ and (b) MAPbBr$_3$ at 10 K and at $\lambda_{ex}$ = 1500 nm. In case of MAPbBr$_3$, the black trace corresponds to the low-energy tail of 2PA-induced PL. Right/left circularly polarized PL($\sigma^-$)/PL($\sigma^+$) under $\sigma^-/\sigma^+$ excitation from (c) MAPbBr$_3$ using resonant 2PA and (d) PMMA-coated MAPbBr$_3$ using typical 1PA at 10 K. For 2PA, we intentionally formed the focus at the interior of the single crystal to make sure bulk excitation.

The surface-induced static effect was further evidenced by zero DOCP under resonant 2PA (Figure 4c), which directly generates cold excitons with well-defined spin orientation in the bulk. Note here that resonant excitation to the exciton line would yield the maximum DOCP in accordance with Figure 3d. Therefore, zero DOCP even under resonant 2PA unambiguously demonstrates that the perovskite crystallizes into a centrosymmetric structure inside the OIHP single crystal, and therefore, static inversion symmetry breaking should originate from surface. However, the origin cannot be simply the boundary surface between the sample and ambient, which is present at all temperatures and even in CsPbBr$_3$. Clearly, our results eliminate the candidates (i), (ii), and (iii). We emphasize that the candidate (ii) is indeed the mechanism for dynamic Rashba effect, which is more prevalent at higher temperatures, whereas static Rashba effect in Figure 3c only occurs at temperatures lower than 90 K. Based upon density functional theory calculations, it was recently suggested that surface reconstruction by the MA cation ordering can give rise to static Rashba effect at low temperatures.[53] This localized inversion symmetry breaking is most significant from the boundary surface to a few layers into the interior of the crystal, where the degree of freedom for the MA cation motion is relatively higher than that in the bulk. In order to prove the feasibility of the model, we prepared a single crystal of MAPbBr$_3$ coated with PMMA. Quite interestingly, we confirmed that the PMMA-coated MAPbBr$_3$ did not show any static Rashba effect (Figure 4d) in a sharp contrast with the pristine crystal of MAPbBr$_3$, yielding a finite DOCP value (Figure 3c). We believe that PMMA passivates the surface defects[55] and constrains MA cations by hydrogen bonding between the cations and the capping polymers.[56] Our observation therefore indicates that the source for static inversion symmetry breaking at the surface is the candidate (iv), surface reconstruction. In fact, low-temperature surface reconstruction was recently observed in MAPbBr$_3$ by real-space imaging.[57] Although PMMA coating can passivate surface defects, we can neglect the impact of surface defects on static Rashba effect simply because of the absence of Rashba splitting in CsPbBr$_3$. In other words, if surface defects are the main mechanism, CsPbBr$_3$ should also exhibit static Rashba effect as halide vacancy is ubiquitous on the surface of perovskites.[15,58] The role of surface recombination and its temperature dependence is more discussed in the Supporting Information (Figures S12 and analysis of TRPL data).

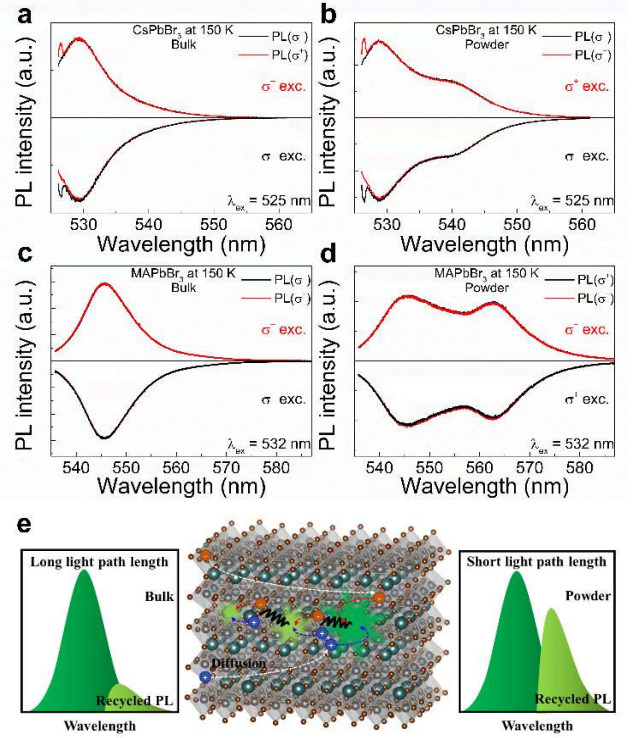

**Figure 5.** Right/left circularly polarized PL($\sigma^-$)/PL($\sigma^+$) under $\sigma^-/\sigma^+$ excitation from (a) bulk CsPbBr$_3$, (b) CsPbBr$_3$ powder, (c) bulk MAPbBr$_3$, and (d) MAPbBr$_3$ powder at 150 K, where DOCP = 0. The indirect transition is more pronounced in powders (smaller in size) due to less efficient photon recycling. (e) Schematic diagram for the photon-recycling effect: the PL undergoes more recycling steps for the longer PL path.

Lastly, we examined the impact of the crystal size on photon recycling in the indirect bandgap. In order to clearly confirm the effect, we carried out polarization-dependent PL spectroscopy on the bulk and powder (~100 μm) samples at 150 K, where static Rashba effect is absent (DOCP = 0). In Figures 5a, 5b, 5c, and 5d, we plot the right/left circularly polarized PL($\sigma^-$)/PL($\sigma^+$) from (a) bulk CsPbBr$_3$, (b) CsPbBr$_3$ powder, (c) bulk MAPbBr$_3$, and (d) MAPbBr$_3$ powder, respectively, at 150 K under $\sigma^-/\sigma^+$ excitation. Our results clearly show that the secondary peak can significantly alter the PL spectrum depending on the degree of photon recycling in both AIHP and OIHP single crystals. Compared with the bulk counterparts, small-sized powders yield the more pronounced peak below the direct gap due to less steps of reabsorption and reemission. This effect is schematically illustrated in Figure 5e.

## CONCLUSION

We have thoroughly investigated both dynamic and static Rashba effects in bromide AIHP and OIHP single crystals. Both perovskite single crystals exhibit dynamic Rashba effect as evidenced by the formation of the indirect gap via thermal distortion of the lattice. While the effect persists in MAPbBr$_3$, dynamic Rashba splitting in CsPbBr$_3$ disappears at low temperatures (< 50 K), which is accompanied by the apparent color change of the crystal. Based on circularly polarized PL spectroscopy, we have also examined static Rashba effect and found that only MAPbBr$_3$ exhibits static Rashba effect with DOCP~3% at 10 K. The static inversion symmetry breaking occurs near the surface, not bulk, as confirmed by nonlinear optical harmonic generation and 2PA. We have identified that the underlying mechanism for static Rashba effect is surface reconstruction of MA cations, which was theoretically predicted and experimentally confirmed recently.[53,57] Considering the surface nature of the effect and a relatively fast spin decoherence time of ~2 ps in bulk perovskite crystals,[52] we believe that time-integrated DOCP of 3% is a rather large value. We have also demonstrated that static Rashba effect in the OIHP can be actively controlled by a simple surface treatment with PMMA, which suppresses the degree of freedom for MA-cation ordering. Finally, the PL spectrum can be significantly affected by photon recycling via the Rashba gap depending on the crystal size. Our results indicate that Rashba effect is less prone to extrinsic defects We believe that our results are important for furthering the halide perovskites into the emerging field of spin-optoelectronics.

## ASSOCIATED CONTENT

**Supporting Information**

## AUTHOR INFORMATION

**Corresponding Author**

**Joon I. Jang** – *Department of Physics, Sogang University, Seoul 04107, South Korea*

*E-mail: jjcoupling@sogang.ac.kr

**Authors**

**Hongsun Ryu** – *Department of Physics, Sogang University, Seoul 04107, South Korea*

**Dae Young Park** – *Department of Energy Science, Sungkyunkwan University, Suwon, 16419, South Korea*

**Kyle M. McCall** – *Department of Chemistry, Northwestern University, Evanston, Illinois 60208, USA*
**Present Address**
Laboratory of Inorganic Chemistry, Department of Chemistry and Applied Biosciences, ETH Zurich, 8093 Zurich, Switzerland
Laboratory for Thin Films and Photovoltaics, Empa - Swiss Federal Laboratories for Materials Science and Technology, CH-8600 Dübendorf, Switzerland

**Hye Ryung Byun** – *Department of Physics, Sogang University, Seoul 04107, South Korea*

**Yongjun Lee** – *Department of Energy Science, Sungkyunkwan University, Suwon, 16419, South Korea*

**Tae Jung Kim** – *Department of Physics, Kyung Hee University, Seoul, 02447, South Korea*

**Mun Seok Jeong** – *Department of Energy Science, Sungkyunkwan University, Suwon, 16419, South Korea*

**Jeongyong Kim** – *Department of Energy Science, Sungkyunkwan University, Suwon, 16419, South Korea*

**Mecouri G. Kanatzidis** – *Department of Chemistry, Northwestern University, Evanston, Illinois 60208, USA*

**Notes**
The authors declare no competing financial interest.
Any additional relevant notes should be placed here.

## ACKNOWLEDGMENT

This work was supported by the Basic Science Research Programs (2017R1D1A1B03035539 and 2020R1F1A1069646) through the National Research Foundation of Korea (NRF), funded by the Korean government. At Northwestern this work was supported by the US Department of Energy, Office of Science, Basic Energy Sciences, under Grant No. SC0012541 (synthesis and structural characterization).